\documentclass[a4paper,12pt,eqno]{article}
\usepackage{amssymb}
\usepackage{theorem}
\newtheorem{theorem}{THEOREM}
\newtheorem{lemma}[theorem]{LEMMA}
\newtheorem{corollary}[theorem]{COROLLARY}
\newtheorem{proposition}[theorem]{PROPOSITION}
\theoremheaderfont{\scshape}

\newcommand{\ket}[1]{\> #1}

\title{{\Large {\bf Limit theorems and absorption problems for one-dimensional correlated random walks}
}}

\author{Norio Konno \\
Department of Applied Mathematics, Yokohama National University \\
79-5 Tokiwadai, Hodogaya-ku, Yokohama, 240-8501, Japan \\
{\it e-mail: konno@ynu.ac.jp}
}

\date{\empty }
\begin{document}
\maketitle

\par\noindent
\begin{small}
{\bf Abstract}. There has recently been considerable interest in quantum walks in connection with quantum computing. The walk can be considered as a quantum version of the so-called correlated random walk. We clarify a strong structural similarity between both walks and study limit theorems and absorption problems for correlated random walks by our PQRS method, which was used in our analysis of quantum walks. 

\footnote[0]{
{\it Abbr. title:} Limit theorems and absorption problems for correlated random walks.
}


\end{small}

\setcounter{equation}{0}
\section{Introduction}

The correlated random walk in one dimension can be defined as follows. We suppose that a particle performs a random walk on a state space $S$. In this paper, we consider three cases as $S$, that is, $S= {\bf Z}, \> S={\bf Z}_+,$ and $S= \{0,1, \ldots , N\},$ where ${\bf Z}$ is the set of integers and ${\bf Z}_+ = \{0, 1, \ldots \}$. The evolution of the correlated random walk is given by 
\begin{eqnarray*} 
&&
P( \mbox{ the particle moves one unit to the left})
\\
&&
\qquad 
=
\left\{
\begin{array}{rl}
p, & \qquad \mbox{if the previous step was to the left,} \\
1-q, & \qquad \mbox{if the previous step was to the right,} 
\end{array}
\right.
\end{eqnarray*}
and
\begin{eqnarray*} 
&&
P( \mbox{ the particle moves one unit to the right})
\\
&&
\qquad 
=
\left\{
\begin{array}{rl}
1-p, & \qquad \mbox{if the previous step was to the left,} \\
q, & \qquad \mbox{if the previous step was to the right.} 
\end{array}
\right.
\end{eqnarray*}
When $p=q$, the probability that the particle moves one unit in the same direction as the last step is $p$, and the probability that the particle moves one unit in the opposite direction as the last step is $1-p$. In particular, if $p=q=1/2$, then the walk is equivalent to the well-known symmetric (non-correlated) random walk, i.e., the particle moves at each step either one unit to the left with probability $1/2$ or one unit to the right with probability $1/2.$ The directions of different steps are independent of each other. 

\par
The correlated random walk has been studied by various authors. Recent work relating to the present paper includes \cite{All2004,AM2001,b2000,b2002,CR1992,CR1994,lb,MS1987,rh,z}. More detailed information on their results will be stated in the next section. 
\par
On the other hand, a discrete-time quantum walk was introduced and intensively studied by \cite{aa,ab} as a quantum variant of the classical random walk and since then considerable work has been done on quantum walks by a number of groups in connection with quantum computing (see \cite{a,ke,t}, for a review). Our group has also been studying quantum walks by using mainly the PQRS method based on combinatorics which is different from Fourier analysis, (see Konno \cite{kqip,k,kqic}, Konno, Namiki and Soshi \cite{kns}, Konno et al. \cite{knss}). Meyer and Blumer \cite{mb} pointed out that the quantum walk can be considered as a quantum version of the correlated random walk. More recently, Romanelli et al. \cite{r} analyzed the quantum walk by separating the quantum evolution into Markovian and interference terms. They reported that if the interference terms are neglected then the resulting equation becomes the Telegraphist's equation which can be considered as a space-time continuum limit of the correlated random walk. It is a well established result for a one-dimensional quantum walk that if (i) it is expressed in terms of pairs of real probabilities and (ii) quantum correlations are neglected, then, it reduces to a two-dimensional Markovian process which is equivalent to a one-dimensional correlated random walk (or, equivalently, to a Telegraphist's equation in the continuum limit). The same fact holds for a one-dimensional correlated (or persistent) random walk: it is equivalent to a Markovian process in two dimensions. As the following definitions show,  there is a strong structural similarity between them. One of our motivations for this study is to clarify the gap between them by using our PQRS method. We present limit theorems and absorption probabilities for correlated random walks as in the case of quantum walks (see \cite{kqic}).
\par
This paper is organized as follows. Section 2 treats the definitions of correlated random walk and the quantum walk, and explains the PQRS method. Moreover we summarize some recent work on correlated random walks. In Section 3, we give two types of limit theorems. Section 4 is devoted to one- and two-boundary absorption problems.

\section{Definition and PQRS Method}
\noindent
First we present the definition of the correlated random walk, which clarifies a connection with the quantum walk. The time evolution of the walk is given by the following transposed transition matrix:
\begin{eqnarray*}
A=
\left[
\begin{array}{cc}
a & b \\
c & d
\end{array}
\right],
\end{eqnarray*}
\par\noindent
where
\begin{eqnarray*}
a=p, \quad b= 1-q, \quad c= 1-p, \quad d= q.
\end{eqnarray*}
In order to avoid trivialities, we assume that $0 <a, d <1$ throughout this paper. In the case of quantum walks, $A$ is a unitary matrix with $a,b,c,d \in {\bf C},$ where ${\bf C}$ is the set of complex numbers. By the unitarity, elements of $A$ satisfy the following relations: $|a|^2 + |c|^2 =|b|^2 + |d|^2 =1, \> a \overline{c} + b \overline{d}=0, \> c= - \triangle \overline{b}, d= \triangle \overline{a}$, where $\overline{z}$ is the complex conjugate of $z \in {\bf C}$ and $\triangle = \det A = ad - bc.$ 
\par
For both correlated random walks and quantum walks, we can think of the evolution as driven by the outcome of tosses of two coins, with each coin determining the evolution at the time step for which that coin is active. The directions R and L can be respectively thought of as corresponding to the heads and tails of the coin, or equivalently as an internal chirality state of the particle. The value of the coin controls the direction in which the particle moves. In fact, when the coin shows L, the particle moves one unit to the left, when it shows R, it moves one unit to the right. Let
\begin{eqnarray*}
\ket{L} = 
\left[
\begin{array}{cc}
1 \\
0  
\end{array}
\right],
\qquad
\ket{R} = 
\left[
\begin{array}{cc}
0 \\
1  
\end{array}
\right],
\end{eqnarray*}
so we have
\begin{eqnarray*}
&& U\ket{L} = a\ket{L} + c\ket{R}, \\
&& U\ket{R} = b\ket{L} + d\ket{R}.
\end{eqnarray*}
\par
From now on, in order to emphasize the connection with the quantum walk, we sometimes use notation taken from this field. Let $\ket{\Psi_{k} (n)} (\in {\bf R}^2)$ be the probability that the correlated random walk is in location $k$ at time $n$, 
\begin{eqnarray*}
\ket{\Psi_{k} (n)} = 
\left[
\begin{array}{cc}
\Psi_{k} ^L (n) \\
\Psi_{k} ^R (n)
\end{array}
\right],
\end{eqnarray*}
with the chirality being left (upper component) or right (lower component). Then the dynamics for $\ket{\Psi_{k} (n)}$ is given by the following transformation:
\begin{eqnarray}
\ket{\Psi_{k} (n+1)} = P \ket{\Psi_{k+1} (n)} + Q \ket{\Psi_{k-1} (n)},
\label{koi}
\end{eqnarray}
where
\begin{eqnarray*}
P= 
\left[
\begin{array}{cc}
a & b \\
0 & 0 
\end{array}
\right], 
\quad
Q=
\left[
\begin{array}{cc}
0 & 0 \\
c & d 
\end{array}
\right].
\end{eqnarray*}
It is noted that $A=P+Q.$ Romanelli et al. \cite{r} gave a similar argument, so we clarify the relation between our setting and theirs. In our notation, the model they considered corresponds to the following correlated random walk:
\begin{eqnarray*}
A= 
\left[
\begin{array}{cc}
\cos^2 \theta & \sin^2 \theta \\
\sin^2 \theta & \cos^2 \theta 
\end{array}
\right],
\end{eqnarray*}
that is, $a=d= \cos^2 \theta$ and $b=c= \sin^2 \theta$ for some value of $\theta$. Therefore, by using Eq.~(\ref{koi}), we have
\begin{eqnarray*}
&& 
\ket{\Psi_{k} ^L (n+1)} 
= 
\cos^2 \theta \ket{\Psi_{k+1} ^L (n)} + \sin^2 \theta \ket{\Psi_{k+1} ^R (n)},
\\
&&
\ket{\Psi_{k} ^R (n+1)} 
= 
\sin^2 \theta \ket{\Psi_{k-1} ^L (n)} + \cos^2 \theta \ket{\Psi_{k-1} ^R (n)}.
\end{eqnarray*}
The above equations can be also obtained when the interference terms are neglected in Eq.~(5) in their paper. We remark that $\Psi_{k} ^j (n)$ (our notation) is equivalent to $P_{k,j} (n)$ (their notation) $(j=L,R).$

One of the essential parts of the present paper is the study on the dependence of some important properties and quantities (e.g., limit distribution, symmetry of distribution, absorption probability) on the initial distribution. Therefore we introduce the collection of initial distributions as follows:
\[
\Phi = \left\{ \varphi =
\left[
\begin{array}{cc}
\alpha \\
\beta   
\end{array}
\right]
:
\alpha + \beta =1, \> \alpha, \beta \ge 0
\right\}.
\]
Let $X_n ^{\varphi}$ be the correlated random walk at time $n$ starting from initial distribution $\varphi \in \Phi$ with $X_{0} ^{\varphi}=0$. Let $\Xi (l,m)$ be the sum over possible paths for which the particle arrives at location $k$ at time $n$ starting from the origin with $l+m=n$ and $-l+m=k$. For example, 
\[
\Xi(3,1)= QPPP+ PQPP+ PPQP+PPPQ.
\]
Then 
\begin{eqnarray*}
P(X_{n} ^{\varphi} = k) 
= [1, 1] \>\> \Xi (l,m) \>
\left[
\begin{array}{cc}
\alpha \\
\beta   
\end{array}
\right],
\end{eqnarray*}
where $\langle {}^t [x,y] | {}^t [u,v] \rangle = xu + yv$, ${\bf 1}={}^t [1, 1],$ $\varphi = {}^t[\alpha, \beta] \in {\bf R}^2$ is an initial distribution with $\alpha + \beta =1, \alpha , \beta \ge 0$, and $t$ means the transposed operator. On the other hand, in the case of quantum walks, 
\[
P(\widetilde{X}_{n} ^{\varphi} =k) = || \Xi(l,m) \varphi ||_2 ^2, 
\]
where $\varphi = {}^t[\alpha, \beta] \in {\bf C}^2$ is the initial qubit state with $|\alpha|^2+|\beta|^2=1$ and $|| {}^t[x, y] ||_p = (|x|^p + |y|^p)^{1/p} \> (p \ge 1)$.

In our treatment of correlated random walks, as well as the matrices $P$ and $Q$, it is convenient to introduce
\[
R=
\left[
\begin{array}{cc}
c & d \\
0 & 0 
\end{array}
\right], 
\quad
S=
\left[
\begin{array}{cc}
0 & 0 \\
a & b 
\end{array}
\right].
\]
When $ad - bc \> (= a+d -1) \not= 0,\>$ $P,Q,R,$ and $S$ form a basis of $M_2 ({\bf R})$ which is the vector space of real $2 \times 2$ matrices. Now, we let
\[
X=
\left[
\begin{array}{cc}
x & y \\
z & w 
\end{array}
\right].
\]
If $ad - bc \not= 0,$ then
\[
X=c_p P + c_q Q + c_r R + c_s S,
\]
where the coefficients $c_p, c_q, c_r, c_s$ are determined by
\[
\left[
\begin{array}{cccc}
c_p \\
c_r \\
c_s \\
c_q 
\end{array}
\right]
=
{1 \over ad-bc} 
\left[
\begin{array}{cccc}
 d & -c &  0 &  0 \\
-b &  a &  0 &  0 \\
 0 &  0 &  d & -c \\
 0 &  0 & -b &  a 
\end{array}
\right]
\left[
\begin{array}{cccc}
x \\
y \\
z \\
w 
\end{array}
\right].
\]
Therefore we can express any $2 \times 2$ matrix $X$ conveniently in the form,
\begin{eqnarray} 
X 
=
{1 \over ad -bc} 
\left\{ (dx-cy) P + (-bz+aw) Q +(-bx+ay) R +(dz-cw) S \right\}. 
\label{eq:pqrs}
\end{eqnarray} 
From now on, we assume that $ad -bc \not= 0.$ We call the analysis based on $P,Q,R,S$ the PQRS method in this paper. 

The $n \times n$ unit and zero matrices are written as $I_n$ and $O_n$ respectively. If $X = I_2$, then Eq.~(\ref{eq:pqrs}) gives
\begin{eqnarray} 
I_2 = {1 \over ad -bc} \left( dP+aQ -bR -cS \right).
\label{eq:identity}
\end{eqnarray} 
The following table of products of $P,Q,R,S$ is very useful in computing some quantities:
\par
\
\par
\begin{center}
\begin{tabular}{c|cccc}
  & $P$ & $Q$ & $R$ & $S$  \\ \hline
$P$ & $aP$ & $bR$ & $aR$ & $bP$  \\
$Q$ & $cS$ & $dQ$& $cQ$ & $dS$ \\
$R$ & $cP$ & $dR$& $cR$ & $dP$ \\
$S$ & $aS$ & $bQ$ & $aQ$ & $bS$ 
\end{tabular}
\end{center}
where $PQ=bR$, for example.

\par
In order to consider absorption problems in Section 4, we describe the absorption probability of a correlated random walk starting from location $k$ on $\{ 0, 1, \ldots , N \}$ ($N \le \infty $) with absorbing boundaries $0$ and $N$ as follows. 

First we consider $N < \infty$ case. Let $\Xi^{(N)} _{k} (n)$ be the sum over possible paths for which the particle first hits 0 at time $n$ starting from $k$ before it arrives at $N$. For example, 
\begin{eqnarray*} 
\Xi^{(3)} _{1} (5) = P^2 Q P Q = ab^2cR.
\end{eqnarray*} 
The probability that the particle first hits 0 at time $n$ starting from $k$ with initial distribution $\varphi (\in \Phi)$ before it arrives at $N$ is
\[
P^{(N)} _{k} (n; \varphi) = || \Xi^{(N)} _k (n) \varphi ||_1.
\]
So the probability that the particle first hits 0 starting from $k$ with initial distribution $\varphi (\in \Phi)$ before it arrives at $N$ is
\[
P^{(N)} _{k} (\varphi) = \sum_{n=0} ^{\infty} P^{(N)} _{k} (n;\varphi).
\]

Next we consider $N = \infty$ case. Let $\Xi^{(\infty)} _{k} (n)$ be the sum over possible paths for which the particle hits 0 at time $n$ starting from $k$. For example,
\begin{eqnarray*} 
\Xi^{(\infty)} _{1} (5) = P^2 Q P Q + P ^3 Q ^2= (a b^2 c + a^2bd) R.
\end{eqnarray*} 
In a similar way, we can define $P^{(\infty)} _{k} (n; \varphi)$ and $P^{(\infty)} _{k} (\varphi)$. 

From now on we review some recent works on correlated random walks. Renshaw and Henderson \cite{rh} considered the $a=d$ case and obtained an expression for $P(X_{n} ^{\varphi} =k)$ and its limiting distribution as $n \to \infty$ with $a \to 1$ for the state space $S={\bf Z}$. Zhang \cite{z} investigated the absorption probability and the expected duration for various types of boundaries such as an absorbing boundary or an elastic boundary. Lal and Bhat \cite{lb} calculated $P(X_{n} ^{\varphi} =k)$ for $S={\bf Z}$, and limiting distribution for $S= \{0, 1, \ldots \}$ and $S= \{0, 1, \ldots N \}$. B\"ohm \cite{b2002} considered the $a=b$ case and studied the distribution of the absorption time and the asymptotic distribution of the maximum for this walk on $S= \{0, 1, \ldots \}$. B\"ohm \cite{b2000} computed $P(X_{n} ^{\varphi} =k)$ for $S={\bf Z}, \> \{0, 1, \ldots \},$ and $\{0, 1, \ldots N \}$ based on Krattenthaler's result (Krattenthaler \cite{kr}) for counting lattice paths with turns. Moreover, when $a = d$, B\"ohm gave the asymptotic behaviour of $P(X_{n} ^{\varphi}=k)$ as $n \to \infty$. In addition, Allaart and Monticino \cite{AM2001} analyzed optimal stopping rules for a general class of correlated random walks. More refined results were obtained by Allaart \cite{All2004}. Chen and Renshaw investigated correlated random walks in higher dimensions \cite{CR1992} and considered a more general walk \cite{CR1994}. Mukherjea and Steele \cite{MS1987} studied a correlated gambler's ruin problem.

\section{Characteristic Function and Limit Theorem}
\noindent
This section treats two different types of limit theorems for correlated random walks $X^{\varphi} _{n}.$ To do so, first we need to know $P(X_{n} ^{\varphi} =k)$ for $n+k=$ even. For fixed $l$ and $m$ with $l+m=n$ and $-l+m=k$, we have
\[
\Xi (l,m)= \sum_{l_i, m_i \ge 0: m_1+ \cdots +m_n=m, l_1+ \cdots +l_n=l} P ^{l_1}Q ^{m_1}P ^{l_2}Q ^{m_2} \cdots P ^{l_n}Q ^{m_n},
\]
since 
\[
P(X_n ^{\varphi} =k) = {}^t {\bf 1} \> \Xi(l,m) \> \varphi.
\]
Noting that $P, Q, R$, and $S$ are a basis of $M_2 ({\bf R})$, $\Xi (l,m)$ has the following form:
\[
\Xi(l,m) = p (l,m) P + q (l,m) Q + r (l,m) R + s (l,m) S.
\]
Our next aim is to obtain explicit forms for $p (l,m), q (l,m), r (l,m),$ and $s (l,m)$. In the case of $n=l+m=4$, we know 
\begin{eqnarray*}
&& \Xi(4,0) = a^3 P, \quad
\Xi(3,1) = 2abc P + a^2b R + a^2c S, \quad \\
&& \Xi(2,2) =  bcd P + abc Q + b(ad+bc) R + c(ad+bc) S, \\
&& \Xi(1,3) = 2bcd Q + bd^2 R + cd^2 S, \quad
\Xi(2,2) = d^3 Q.
\end{eqnarray*}
So, for example,
\[
p (3,1)=2abc, \quad q (3,1)=0, \quad r (3,1)=a^2b, \quad s (3,1)=a^2c.
\]
\par
Now the following key lemma can be obtained by using our combinatorial method. 
\begin{lemma} 
\label{lem1}
We consider correlated random walks with $abcd \not= 0$. Suppose that $l,m \ge 0$ with $l+m=n$, then we have 
\par\noindent
\hbox{(i)} for $l \wedge m (= \min \{l,m \}) \ge 1$,  
\begin{eqnarray*}
\Xi(l,m) 
= a^l d^m 
\sum_{\gamma =1} ^{l \wedge m} 
\left({bc \over ad} \right)^{\gamma}
{l-1 \choose \gamma- 1} 
{m-1 \choose \gamma- 1} 
\times 
\Biggl[ {l- \gamma \over a \gamma } P + {m - \gamma \over d \gamma} Q + {1 \over c} R + {1 \over b} S 
\Biggr].
\end{eqnarray*}
\par\noindent
\hbox{(ii)} for $l (=n) \ge 1, m = 0$,  
\[
\Xi(l,0) = a^{l-1} P.
\]
\par\noindent
\hbox{(iii)} for $l = 0, m (=n) \ge 1$,  
\[
\Xi(0,m) = d^{m-1} Q.
\]
\end{lemma} 
\noindent
The proofs of parts (ii) and (iii) are trivial. The proof of part (i) is based on enumerating the paths of drift $l+m=n$. To do so, we consider the following 4 cases:
\begin{eqnarray*}
&&
p(l,m): \quad
\overbrace{PP \cdots P}^{w_1} \overbrace{QQ \cdots Q}^{w_2} \overbrace{PP \cdots P}^{w_3} \cdots \overbrace{QQ \cdots Q}^{w_{2 \gamma}} \overbrace{PP \cdots P}^{w_{2 \gamma+1}}, \\
&&
q(l,m): \quad
\overbrace{QQ \cdots Q}^{w_1} \overbrace{PP \cdots P}^{w_2} \overbrace{QQ \cdots Q}^{w_3} \cdots \overbrace{PP \cdots P}^{w_{2 \gamma}} \overbrace{QQ \cdots Q}^{w_{2 \gamma+1}}, \\
&&
r(l,m): \quad
\overbrace{PP \cdots P}^{w_1} \overbrace{QQ \cdots Q}^{w_2} \overbrace{PP \cdots P}^{w_3} \cdots \overbrace{QQ \cdots Q}^{w_{2 \gamma}}, \\
&&
s(l,m): \quad
\overbrace{QQ \cdots Q}^{w_1} \overbrace{PP \cdots P}^{w_2} \overbrace{QQ \cdots Q}^{w_3} \cdots \overbrace{PP \cdots P}^{w_{2 \gamma}},
\end{eqnarray*}
where $w_1, w_2, \ldots , w_{2 \gamma+1} \ge 1$ and $\gamma \ge 1$. We remark that $p(l,m), q(l,m), r(l,m), s(l,m)$ in $\Xi(l,m)$ correspond to the types of paths described above in terms of $P$ and $Q$, respectively. A similar argument can be seen in the proof of Theorem 2.1 of B\"ohm \cite{b2000}. On the other hand, for the quantum walk case, this proof appears in Konno \cite{k}. One of the main difference between the correlated random walk and quantum walk is the sign of the term $(bc/ad)$. That is, 
\begin{eqnarray*} 
{bc \over ad}
=
\left\{
\begin{array}{rl}
(1-a)(1-d)/ad > 0, 
& \quad \mbox{correlated random walk,} \\
- (1-|d|^2)/|a|^2 < 0, 
& \quad \mbox{quantum walk.} 
\end{array}
\right.
\end{eqnarray*}
The negativity in the quantum walk case corresponds to the interference effect. For the convenience of readers, we give a outline of the proof based on \cite{k}.

\noindent
{\em Proof.} Here we consider only $p(l,m)$ case. Since the proofs of the other cases are similar, we will omit them. First we assume $l \ge 2$ and $m \ge 1$. In order to compute $p (l,m)$, it is sufficient to consider only the following case:
\begin{eqnarray*}
C(P,w)^{(2 \gamma +1)} _n (l,m) = \overbrace{PP \cdots P}^{w_1} \overbrace{QQ \cdots Q}^{w_2} \overbrace{PP \cdots P}^{w_3} \cdots \overbrace{QQ \cdots Q}^{w_{2 \gamma}} \overbrace{PP \cdots P}^{w_{2 \gamma+1}}, 
\end{eqnarray*}
where $w=(w_1, w_2, \ldots , w_{2 \gamma+1}) \in {\bf Z}_+ ^{2 \gamma +1}$ with $w_1, w_2, \ldots , w_{2 \gamma+1} \ge 1$ and $\gamma \ge 1$. For example, the string $PQP$ has $w_1=w_2=w_3=1$ and $\gamma =1$. We should remark that $l$ is the number of $P$s and $m$ is the number of $Q$s, so we have
\begin{eqnarray*}
&& l= w_1 + w_3 + \cdots + w_{2 \gamma +1}, \\
&& m= w_2 + w_4 + \cdots + w_{2 \gamma}.
\end{eqnarray*}
Moreover $2 \gamma +1$ is the number of clusters of $P$'s and $Q$'s. Next we consider the range of $\gamma$. The minimum is $\gamma =1$, that is, 3 clusters. This case is
\[
P \cdots P Q \cdots Q P \cdots P. 
\]
The maximum is $\gamma = (l-1) \wedge m$. This case includes the patterns:
\[
PQPQPQ \cdots PQPQ P \cdots P \> (l-1 \ge m), \quad PQPQPQ \cdots PQ P Q \cdots Q P \> (l-1 \leq m),
\]
for examples. Here we introduce a set of sequences with $2 \gamma +1$ components: for fixed $\gamma \in [1, (l-1) \wedge m]$, 
\begin{eqnarray*}
W(P,2 \gamma +1) 
&=& \{ w = (w_1, w_2, \cdots , w_{2 \gamma +1}) \in {\bf Z} ^{2 \gamma +1} : 
w_1 + w_3 + \cdots + w_{2 \gamma +1} =l, \\ 
&& \qquad \qquad  w_2 + w_4 + \cdots + w_{2 \gamma} =m, \> w_1, w_2, \ldots, w_{2\gamma}, w_{2 \gamma +1} \ge 1 \}. 
\end{eqnarray*}
From the following relations:
\begin{eqnarray*}
P^2 = aP, \quad Q^2 = dQ, \quad PQ=bR, \quad R^2= cR, \quad RP=cP,
\end{eqnarray*}
we have
\begin{eqnarray*}
C(P,w)_n ^{(2 \gamma +1)}(l,m) 
&=& a^{w_1-1} P d^{w_2 -1} Q a^{w_3-1} P \cdots d^{w_{2 \gamma} -1} Q a^{w_{2 \gamma +1} -1} P \\
&=& a^{l-(\gamma +1)} d^{m - \gamma} (PQ)^{\gamma} P \\
&=& a^{l-(\gamma +1)} d^{m - \gamma} b^{\gamma} R^{\gamma} P \\ 
&=& a^{l-(\gamma +1)} d^{m - \gamma} b^{\gamma} c^{\gamma-1} RP \\ 
&=& a^{l-(\gamma +1)} d^{m - \gamma} b^{\gamma} c^{\gamma} P,
\end{eqnarray*}
where $w \in W(P, 2 \gamma +1)$. For $l \ge 2, m \ge 1$, that is, $\gamma \ge 1$, we obtain
\begin{eqnarray*}
C(P,w)_n ^{(2 \gamma +1)}(l,m) = 
a^{l-(\gamma +1)}  b^{\gamma} c^{\gamma} d^{m - \gamma} P.
\end{eqnarray*}
Note that the right hand side of the above equation does not depend on $w \in W(P, 2 \gamma +1)$. Finally we compute the number of $w=(w_1,w_2, \ldots, w_{2\gamma}, w_{2 \gamma +1})$ satisfying $w \in  W(P, 2 \gamma +1)$ by a standard combinatorial argument as follows:
\[
|W(P, 2 \gamma +1)| = { l-1 \choose \gamma } { m-1 \choose \gamma -1}.
\]
From the above observation, we obtain
\begin{eqnarray*}
p (l,m) P
&=& \sum_{\gamma =1} ^{(l-1) \wedge m}
\sum_{w \in W(P,2 \gamma +1)}
C(P,w)_n ^{(2 \gamma +1)}(l,m) \\
&=&
\sum_{\gamma =1} ^{(l-1) \wedge m}
|W(P,2 \gamma +1)| C(P,w)_n ^{(2 \gamma +1)}(l,m) \\
&=&
\sum_{\gamma =1} ^{(l-1) \wedge m}
{ l-1 \choose \gamma } { m-1 \choose \gamma -1} 
a^{l-(\gamma +1)}  b^{\gamma} c^{\gamma} d^{m - \gamma} P.
\end{eqnarray*}
So we conclude that
\begin{eqnarray*}
p (l,m) =
\sum_{\gamma =1} ^{(l-1) \wedge m} { l-1 \choose \gamma } { m-1 \choose \gamma -1}a^{l-(\gamma +1)}  b^{\gamma} c^{\gamma} d^{m - \gamma}.
\end{eqnarray*}
When $l \ge 1$ and $m=0$, it is easy to see that
\begin{eqnarray*}
p(l,0) P = P^l = a^{l-1} P.
\end{eqnarray*}
Furthermore, when $l=1, m \ge 1$ and $l = 0, m \ge 0$, it is clear that
\begin{eqnarray*}
p (l,m) = 0.
\end{eqnarray*}
Similarly we have the following results for $q(l,m), r(l,m)$ and $s(l,m)$: 
if $l \ge 1$ and $m \ge 2$, then
\begin{eqnarray*}
q (l,m) =
\sum_{\gamma =1} ^{l \wedge (m-1)} { l-1 \choose \gamma -1} { m-1 \choose \gamma } a^{l-\gamma}  b^{\gamma} c^{\gamma} d^{m - (\gamma +1)},
\end{eqnarray*}
if $l = 0$ and $m \ge 1$, then $q (0,m) = d^{m-1}$. When $l \ge 1, m = 1$ and $l \ge 0, m=0$, we know $q (l,m) = 0$. If $l,m \ge 1$, then
\begin{eqnarray*}
r (l,m) =
\sum_{\gamma =1} ^{l \wedge m} { l-1 \choose \gamma -1} { m-1 \choose \gamma -1}a^{l-\gamma}  b^{\gamma} c^{\gamma -1} d^{m - \gamma},
\end{eqnarray*}
if $l \wedge m =0$, then $r (l,m) = 0$. If $l,m \ge 1$, then
\begin{eqnarray*}
s (l,m) =
\sum_{\gamma =1} ^{l \wedge m} { l-1 \choose \gamma -1} { m-1 \choose \gamma -1}a^{l-\gamma}  b^{\gamma -1} c^{\gamma} d^{m - \gamma},
\end{eqnarray*}
if $l \wedge m =0$, then $s (l,m) = 0$. Therefore, when $l \wedge m \ge 1$, we have
\begin{eqnarray*}
\Xi (l,m) 
=
a^l d^m \sum_{\gamma =1} ^{l \wedge m} \left( {bc \over ad} \right)^{\gamma}
{ l-1 \choose \gamma -1} { m-1 \choose \gamma -1}
\left[ {l- \gamma \over a \gamma } P + {m - \gamma \over d \gamma} Q + {1 \over c} R + {1 \over b} S \right].
\end{eqnarray*}
So the proof of Lemma \ref{lem1} is complete. 
 
\par
\
\par
In order to obtain an expression for the distribution of $X_n ^{\varphi}$, we will use the following facts: for $k=1,2, \ldots , [n/2],$
\begin{eqnarray*}
&& 
\sum_{\gamma =1} ^{k}
\left({bc \over ad} \right)^{\gamma -1}
{1 \over \gamma} 
{k-1 \choose \gamma- 1}  
{n-k-1 \choose \gamma- 1} 
=
{}_2F_1(-(k-1), -\{(n-k)-1\}; 2 ;bc/ad),
\\
&&
\sum_{\gamma =1} ^{k}
\left({bc \over ad} \right)^{\gamma -1}
{k-1 \choose \gamma- 1}  
{n-k-1 \choose \gamma- 1} 
=
{}_2F_1(-(k-1), -\{(n-k)-1\}; 1 ;bc/ad),
\end{eqnarray*}
where $[x]$ is the integer part of $x$ and ${}_2F_1 (a, b; c ;z)$ is the hypergeometric series. In general, as for orthogonal polynomials, see Andrews, Askey and Roy \cite{aar}, for example. Then the distribution of $X^{\varphi} _n$ can be derived from Lemma \ref{lem1} as follows.

\begin{lemma} 
\label{lem2}
For $n \ge 2$ and $k=1,2, \ldots , [n/2],$ 
\begin{eqnarray*}
&&
P(X^{\varphi} _n=n-2k) \\
&&
=  a^{k-2} d^{n-k-2}
\Biggl[
bc
\Bigl\{
(dk+c(n-k)) a \alpha + (bk + a(n-k)) d \beta  
\Bigr\} F_2^{(n,k)} 
\\
&&
\qquad \qquad \qquad \qquad \qquad \qquad \qquad \qquad \qquad \qquad \qquad
+(ac \alpha + bd \beta) \triangle F_1 ^{(n,k)}
\Biggr],
\\
&&
P(X^{\varphi} _n=-(n-2k)) 
\\
&&
=  a^{k-2} d^{n-k-2}
\Biggl[ 
bc
\Bigl\{
(ck+ d(n-k)) a \alpha + (ak + b(n-k)) d \beta  
\Bigr\} F_2^{(n,k)}
\\
&&
\qquad \qquad \qquad \qquad \qquad \qquad \qquad \qquad \qquad \qquad \qquad
+ (ac \alpha + bd \beta) \triangle F_1 ^{(n,k)}
\Biggr],
\end{eqnarray*}
for $n \ge 1$, 
\begin{eqnarray*}
P(X^{\varphi} _n=n) = d^{n-1} (c \alpha + d \beta), \qquad 
P(X^{\varphi} _n=-n) = a^{n-1} (a \alpha + b \beta),
\end{eqnarray*}
where $\triangle = ad -bc,$ and $F_i ^{(n,k)}= {}_2 F_1(-(k-1),-\{(n-k)-1\};i;bc/ad) \> (i=1,2)$.
\end{lemma}

The above result for $\varphi ={}^t [0,1]$ and $\varphi = {}^t [1,0]$ corresponds to Eqs.~(19) and (20) in Corollary 2.6 of \cite{b2000}, respectively. By using Lemma \ref{lem2}, we have an expression for the characteristic function of $X^{\varphi} _n$. This result will be used in order to obtain limit theorems of $X^{\varphi} _n$.

\begin{theorem} 
\label{th3}
When $abcd \not= 0$, we have
\begin{eqnarray*}
E(e^{i \xi X_n ^{\varphi}}) 
&=& 
\Biggl[ 
a^{n-1}(a \alpha + b \beta) + d^{n-1} (c \alpha + d \beta) 
\Biggr]
\cos (n \xi) 
\\
&&
+ i
\Biggl[ 
- a^{n-1}(a \alpha + b \beta) + d^{n-1} (c \alpha + d \beta) 
\Biggr]
\sin (n \xi)
\\
&& 
+ \sum_{k=1}^{\left[{n-1 \over 2}\right]}
a^{k-2}d^{n-k-2}
\Biggl[ 
\Bigl[
bcn 
\Bigl\{
(c+d)a \alpha + (a+b)d \beta
\Bigr\} F_2 ^{(n,k)}
\\
&&
\qquad \qquad \qquad \qquad \qquad \qquad \qquad 
+ 2(ac \alpha + bd \beta) \triangle F_1 ^{(n,k)}
\Bigr]
\cos ((n-2k) \xi) 
\\
&&
\qquad \qquad 
+ ibc (n-2k) 
\Bigl\{
(c-d)a \alpha + (a-b) d \beta
\Bigr\}
F_2 ^{(n.k)}
\sin ((n-2k) \xi)
\Biggr] 
\\
&& 
+
I \Biggl( {n \over 2}-\Biggl[{n \over 2} \Biggr],0 \Biggr)
(ad)^{n/2-2}
\\
&&
\qquad \qquad 
\times
\left[
{bcn \over 2} 
\left\{
(c+d)a \alpha + (a+b)d \beta
\right\} F_2 ^{(n,k)} + 
(ac \alpha + bd \beta) \triangle F_1 ^{(n,k)}
\right],
\end{eqnarray*}
where $I(x,0)=1 (resp. =0)$ if $x=0$ (resp. $x \not= 0$). 
\end{theorem}

\noindent
From this theorem, we have the $m$th moment of $X^{\varphi} _n$ in the standard fashion. The following result can be used in order to study symmetry of distribution of $X^{\varphi} _n$.  

\begin{corollary} 
\label{cor4}
We assume that $abcd \not= 0$. 
\par\noindent
(i) When $m$ is odd, we have
\begin{eqnarray*}
&&
E((X_n ^{\varphi}) ^m) 
= 
\Biggl[ 
- a^{n-1}(a \alpha + b \beta) + d^{n-1} (c \alpha + d \beta) 
\Biggr] n^m
\\
&&
\qquad \qquad \qquad \qquad 
+ \sum_{k=1}^{\left[{n-1 \over 2}\right]}
a^{k-2}d^{n-k-2}
bc (n-2k)^{m+1} 
\Bigl\{
(c-d)a \alpha + (a-b)d \beta
\Bigr\} F_2^{(n,k)}.
\end{eqnarray*}
\par\noindent
\hbox{(ii)} When $m$ is even, we have
\begin{eqnarray*}
&&
E((X_n ^{\varphi}) ^m) 
= 
\Biggl[ 
a^{n-1}(a \alpha + b \beta) + d^{n-1} (c \alpha + d \beta) 
\Biggr] n^m
\\
&&
\qquad \qquad
+ \sum_{k=1}^{\left[{n-1 \over 2}\right]}
a^{k-2}d^{n-k-2}
(n-2k)^m
\\
&&
\qquad \qquad \qquad \qquad 
\times
\Biggl[ 
bcn 
\Bigl\{
(c+d)a \alpha + (a+b)d \beta
\Bigr\} F_2 ^{(n,k)}
+ 2(ac \alpha + bd \beta) \triangle F_1 ^{(n,k)}
\Biggr].
\end{eqnarray*}
\end{corollary}

\noindent
By using Lemma \ref{lem2} and Corollay \ref{cor4} (1) with $m=1$, we obtain the following necessary and sufficient condition for symmetry of distribution of $X_n ^{\varphi}$ with $a=d$. 

\begin{proposition}
\label{pro5}
\par\noindent
(i) If $a=d \in (0,1)$ and $a \neq 1/2$, then 
\[
\Phi_{s} = \Phi_0 = \{ \varphi = {}^t [1/2, 1/2] \},
\]
(ii) if $a=d=1/2$, then 
\[
\Phi_{s} = \Phi_0 = \Phi,
\]
where
\begin{eqnarray*}
\Phi_s &=&  \{ \varphi \in 
\Phi : \> 
P(X_n ^{\varphi}=k) = P(X_n ^{\varphi}=-k) \>\> 
\hbox{for any} \> n \in {\bf Z}_+ \> \hbox{and} \> k \in {\bf Z}
\}, 
\\
\Phi_0 &=& \left\{ \varphi \in 
\Phi : \> 
E(X_n ^{\varphi})=0 \>\> \hbox{for any} \> n \in {\bf Z}_+
\right\}.
\end{eqnarray*}
\end{proposition}

\noindent
The dependence of the symmetry of the distribution on the initial condition is remarkably different between classical and quantum walks. For results on the symmetry of the quantum walk, see \cite{kqip,k,kqic,kns}. In the last part of this section, we will treat two types of limit theorem for $X_n ^{\varphi}$ with different scalings. The first limit theorem for the case of $X_n ^{\varphi}/ \sqrt{n}$ shows that the correlated random walk considered here behaves like as a classical (non-correlated) random walk. The following result can be derived from a consequence of Theorem 3.1 in B\"ohm \cite{b2000} with $m, r, t \to 0, \> \alpha \to a$ and $z \to (1-a)/a$.

\begin{theorem} 
\label{th6}
Consider the correlated random walk with 
\begin{eqnarray*}
A=
\left[
\begin{array}{cc}
a & 1-a \\
1- a & a
\end{array}
\right].
\end{eqnarray*}
If $n \to \infty$, then 
\[
{X_n ^{\varphi} \over \sqrt{n}} \quad \Rightarrow \quad W^{\varphi},
\]
where the distribution of $W^{\varphi}$ is $N(0, a/(1-a))$, with $N(m, \sigma^2)$ the normal distribution with mean $m$ and variance $\sigma^2$ and $Y_n \Rightarrow Y$ means that $Y_n$ converges weakly to a limit $Y$.
\end{theorem}

\noindent\par
It is noted that the above limit distribution is independent of the initial distribution $\varphi \in \Phi$. Theorem \ref{th6} corresponds to Eq.~(16) in \cite{r}, since variance $a/(1-a)$ becomes $\cos^2 \theta/(1-\cos^2 \theta) = \cot^2 \theta = D( \theta )$ in their setting.

Next we consider how we renormalize $X_n ^{\varphi}$ when $a=a_n \to 1$, that is, the above variance $a_n/(1-a_n) \to \infty$. The answer for this question is the following theorem.

\begin{theorem} 
\label{th7}
Consider the correlated random walk with 
\begin{eqnarray*}
A_n=
\left[
\begin{array}{cc}
a_n & 1-a_n \\
1- a_n & a_n
\end{array}
\right],
\end{eqnarray*}
where 
\begin{eqnarray}
a_n = 1 - {\theta \over n},
\end{eqnarray}
and a fixed $\theta \in (0,1)$. If $n \to \infty$, then 
\[
{X_n ^{\varphi} \over n} \quad \Rightarrow \quad Z^{\varphi},
\]
where the probability measure of $Z^{\varphi}$ is the sum of its atomic part 
$\mu_1$ and the absolutely continuous part $\mu_2$ as follows. The atomic part 
is 
\begin{eqnarray*}
\mu_1 = e^{- \theta} (\alpha \delta_{-1} + \beta \delta_{1}),
\end{eqnarray*}
where $\delta_x$ is the Dirac measure at location $x$. The absolutely continuous part $\mu_2$ is given by the density function:
\[
f(x)
= \frac{\theta e^{- \theta}}{2} 
\biggl[ I_0 ( \theta \sqrt{1-x^2}) 
+ {1 \over \sqrt{1-x^2}} I_1 ( \theta \sqrt{1-x^2})
\biggr],
\]
for $x \in (- 1, 1)$, where $I_{\nu} (z)$ is the modified Bessel function of the order $\nu$, that is, 
\[
I_{\nu} (z) 
= \sum_{n=0} ^{\infty} {(z/2)^{\nu +2n} \over n ! \> \Gamma (\nu + n + 1)},
\]
and $\Gamma (z)$ is the gamma function.
\end{theorem}
\par\noindent
{\em Proof.} We begin with an asymptotic result on the hypergeometric series:
\begin{lemma} 
\label{lem8}
Let $a_n = 1 - (\theta/n)$ with $0< \theta <1$.  
If $n \to \infty$ with $k/n=x \in (0, 1/2)$, then
\begin{eqnarray*}
&& 
{}_2 F_1(-(k-1),-\{(n-k)-1\};1;(1-a_n)^2/a_n ^2) 
\to I_0 (2 \theta \sqrt{1-x^2}),
\\
&&
{}_2 F_1(-(k-1),-\{(n-k)-1\};2;(1-a_n)^2/a_n ^2) 
\to {1 \over \theta \sqrt{x(1-x)}} I_1 (2 \theta \sqrt{1-x^2}).
\end{eqnarray*}
\end{lemma}
One of the above derivations can be seen on page 411 of \cite{rh}. By using Theorem \ref{th3} and Lemma \ref{lem8}, we obtain the following asymptotics of characteristic function $E(e^{i \xi X^{\varphi}_n/n})$: if $n \to \infty$ with $k/n=x \in (0,1/2)$, then
\begin{eqnarray*}
&& 
E(e^{i \xi {X_n ^{\varphi} \over n}}) 
\quad \to \quad  
\\
&&
e^{- \theta} \cos \xi
+ i e^{- \theta} (- \alpha + \beta) \sin \xi
\\
&&
\qquad 
+
\theta e^{- \theta} 
\int_0 ^{1/2}
\biggl[ 2 I_0 ( 2 \theta \sqrt{x(1-x)}) 
+ {1 \over \sqrt{x(1-x)}} I_1 ( 2 \theta \sqrt{x(1-x)})
\biggr]
\cos ((1-2x) \xi) dx.
\end{eqnarray*}
Therefore we have
\begin{eqnarray*}
&& 
\lim_{n \to \infty} E(e^{i \xi {X_n ^{\varphi} \over n}}) 
=
e^{- \theta} \cos \xi
+ i e^{- \theta} (- \alpha + \beta) \sin \xi
\\
&&
\qquad 
+
\frac{\theta e^{- \theta}}{2} 
\int_{-1} ^{1} 
\biggl[ I_0 ( \theta \sqrt{1-x^2}) 
+ {1 \over \sqrt{1-x^2}} I_1 ( \theta \sqrt{1-x^2})
\biggr]
\cos (x \xi)
dx
\\
&&
\qquad \qquad 
=
e^{- \theta}
\int_{-1} ^{1}
\Biggl[
(\alpha \delta_{-1}(x) + \beta \delta_1(x))
\\
&&
\qquad \qquad \qquad \qquad \qquad
+ \frac{\theta}{2} 
\biggl\{
I_0 ( \theta \sqrt{1-x^2}) 
+ {1 \over \sqrt{1-x^2}} I_1 ( \theta \sqrt{1-x^2})
\biggr\}
\Biggr]
e^{i \xi x}
dx,
\end{eqnarray*}
where
\[
\int g(x) \delta_a (x) dx = g(a).
\]
So the desired conclusion is obtained. 
\par
\
\par
Interestingly, Theorem \ref{th7} shows that the limit distribution is a mixture of quantum and classical parts. That is, the atomic part $\mu_1$, having two peaks, corresponds to a quantum walk (as for the limit theorem for the quantum walk, see \cite{kqip,k}) and the absolutely continuous part $\mu_2$, having a bell shape like a normal distribution, corresponds to a classical random walk.

\section{Absorption Problem}
\noindent
In this section we consider absorption problems for correlated random walk on state spaces $\{0,1, \ldots \}$ or $\{0,1, \ldots, N\}$ by using the PQRS method as in the case of the quantum walk (see \cite{knss}).

Before we move to the correlated random walk, first we describe the classical (non-correlated) random walk on a finite set $\{0,1, \ldots, N\}$ with two absorption barriers at locations $0$ and $N$ (see \cite{d,gs}, for examples). The particle moves at each step either one unit to the left with probability $p$ or one unit to the right with probability $q=1-p$ until it hits one of the absorption barriers. The directions of different steps are independent of each other. The random walk starting from $k \in \{0,1, \ldots, N \}$ at time $n$ is denoted by $Y^k _n$ here. Let 
\[
T_m = \min \{ n \ge 0 : Y^k _n = m \}
\]
be the time of the first visit to $m \in \{0,1, \ldots, N \}$. Using the subscript $k$ to indicate $Y^k _0=k$, we let
\[
P^{(N)} _k = P_k (T_0 < T_N)
\]
be the probability that the particle hits $0$ starting from $k$ before it arrives at $N$. The absorption problem is also known as the Gambler's ruin problem. To obtain $P^{(N)} _k$, we use the difference equation
\begin{eqnarray}
P^{(N)} _k = p P^{(N)} _{k-1} + q P^{(N)} _{k+1},  
\label{eq:def}
\end{eqnarray}
for $1 \leq k \leq N-1$ with boundary conditions:
\begin{eqnarray}
P^{(N)} _0=1, \quad  P^{(N)} _N=0.  
\label{eq:bc}
\end{eqnarray}
Our basic strategy is to consider a similar equation in the correlated random walk case and to apply our PQRS method to it.

From now on we focus on $1 \leq k \leq N-1$ case for the correlated random walk. So we consider $n \ge 1$ case. Noting that $\{ P,Q,R,S \}$ is a basis of $M_2 ({\bf R})$, $\Xi^{(N)} _{k} (n)$ can be written as
\begin{eqnarray*} 
\Xi^{(N)} _{k} (n) = p^{(N)} _{k} (n) P + q^{(N)} _{k} (n) Q + r^{(N)} _{k} (n) R + s^{(N)} _{k} (n) S.
\end{eqnarray*} 
From the definition of $\Xi^{(N)} _{k} (n)$, it is easily shown that there exist only two types of paths, that is, $P \to \cdots \to P$ and $Q \to \cdots \to P$, since we consider only a hitting time to $0$ before it arrives at $N$. Therefore we see that $q^{(N)} _{k} (n)=s^{(N)} _{k} (n)=0 \> (n \ge 1)$.

We assume $N \ge 3$. Noting that the definition of $\Xi^{(N)} _{k} (n)$, we have 
\begin{eqnarray*} 
\Xi^{(N)} _{k} (n) = \Xi^{(N)} _{k-1} (n-1)P + \Xi^{(N)} _{k+1} (n-1)Q, 
\end{eqnarray*} 
for $1 \leq k \leq N-1$. The above equation corresponds to the difference equation, i.e., Eq.~(\ref{eq:def}) for the classical random walk. A similar approach can be seen in \cite{lb}. Then we have
\begin{eqnarray*} 
&& p^{(N)} _{k} (n) = a p^{(N)} _{k-1} (n-1)+ c r^{(N)} _{k-1} (n-1), \\ 
&& r^{(N)} _{k} (n) = b p^{(N)} _{k+1} (n-1)+ d r^{(N)} _{k+1} (n-1).
\end{eqnarray*} 
Next we consider boundary conditions related to Eq.~(\ref{eq:bc}) as in the case of the classical random walk. When $k=N$, 
\[
P^{(N)} _{N} (0;\varphi) = {}^t {\bf 1} \> \Xi^{(N)} _{N} (0) \> \varphi = 0,
\]
for any $\varphi \in \Phi$. So we take $\Xi^{(N)} _{N} (0) = O_2$. In this case, Eq.~(\ref{eq:pqrs}) gives   
\[
p^{(N)} _{N} (0) =  r^{(N)} _{N} (0) = 0.
\]
If $k=0$, then
\[
P^{(N)} _{0} (0;\varphi) =  {}^t {\bf 1} \> \Xi^{(N)} _{0} (0) \> \varphi = 1,
\]
for any $\varphi \in \Phi$. So we choose $\Xi^{(N)} _{0} (0) = I_2$. From Eq.~(\ref{eq:identity}), we have
\[
p^{(N)} _{0} (0) = {d \over ad -bc}, 
\>\> r^{(N)} _{0} (0) = {-b \over ad -bc}.
\]
Let
\begin{eqnarray*} 
v^{(N)} _{k} (n) = 
\left[
\begin{array}{cc}
p^{(N)} _{k} (n) \\
r^{(N)} _{k} (n)   
\end{array}
\right]. 
\end{eqnarray*}
Therefore we can formulate the absorption problems as follows: 
for $n \ge 1$ and $1 \leq k \leq N-1$, 
\begin{eqnarray} 
&& v^{(N)} _{k} (n) = 
\left[
\begin{array}{cc}
a & c \\
0 & 0 
\end{array}
\right] 
v^{(N)} _{k-1} (n-1)
+
\left[
\begin{array}{cc}
0 & 0 \\
b & d 
\end{array}
\right] 
v^{(N)} _{k+1} (n-1),
\label{nana}
\end{eqnarray}
and for $1 \leq k \leq N$, 
\begin{eqnarray*}
&& v^{(N)} _{0} (0) 
= 
{1 \over ad -bc }
\left[
\begin{array}{cc}
d  \\
-b
\end{array}
\right] 
, \quad
v^{(N)} _{k} (0) 
= 
\left[
\begin{array}{cc}
0  \\
0
\end{array}
\right].
\end{eqnarray*}
Moreover,
\begin{eqnarray*}
v^{(N)} _{0} (n) 
=
v^{(N)} _{N} (n) 
= 
\left[
\begin{array}{cc}
0  \\
0
\end{array}
\right], 
\end{eqnarray*}
for $n \ge 1$. So the definition of $P^{(N)} _{k} (\varphi)$ gives 
\begin{eqnarray} 
P^{(N)} _{k} (\varphi) = \sum_{n=1} ^{\infty} P^{(N)} _{k} 
(n; \varphi),
\label{hati}
\end{eqnarray}
where $\varphi = {}^t[\alpha, \beta] \in \Phi$ and 
\begin{eqnarray}
P^{(N)} _{k} (n; \varphi) 
= (a \alpha + b \beta) p^{(N)} _k (n) + (c \alpha + d \beta) r^{(N)} _k (n). 
\label{kyu}
\end{eqnarray}
To solve $P^{(N)} _{k} (\varphi)$, we introduce generating functions of $p^{(N)} _k (n)$ and $r^{(N)} _k (n)$ as follows:
\begin{eqnarray*}
\widetilde{p}^{(N)} _k (z) = \sum_{n=1} ^{\infty} p^{(N)} _k (n) z^n,
\qquad  \widetilde{r}^{(N)} _k (z) = \sum_{n=1} ^{\infty} r^{(N)} _k (n) z^n.
\end{eqnarray*}
From Eqs.~(\ref{hati}) and (\ref{kyu}), we obtain the following lemma:

\begin{lemma} 
\label{lem9}
We assume that $0 < a, d <1$ with $c=1-a, b=1-d$ and $ad -bc \not = 0.$ Then we have
\begin{eqnarray}
P^{(N)} _{k} (n; \varphi) 
= (a \alpha + b \beta) \widetilde{p}^{(N)} _k (1) 
+ (c \alpha + d \beta) \widetilde{r}^{(N)} _k (1). 
\label{jyu}
\end{eqnarray}
\end{lemma}
\noindent
By Eq.~(\ref{nana}), we have
\begin{eqnarray} 
&& \widetilde{p}^{(N)} _k (z) = a z \widetilde{p}^{(N)} _{k-1} (z) + cz \widetilde{r}^{(N)} _{k-1} (z), 
\label{jyuiti}
\\
&& \widetilde{r}^{(N)} _k (z) = b z \widetilde{p}^{(N)} _{k+1} (z) + dz \widetilde{r}^{(N)} _{k+1} (z). 
\label{eq:req}
\end{eqnarray} 
To solve these, we remark that both $\widetilde{p}^{(N)} _k (z)$ and $\widetilde{r}^{(N)} _k (z)$ satisfy the same recurrence:   
\begin{eqnarray*} 
&& 
d \widetilde{p}^{(N)} _{k+2} (z) - \left( \triangle z+{1 \over z} \right) \widetilde{p}^{(N)} _{k+1} (z) + a \widetilde{p}^{(N)} _{k} (z) = 0,
\\
&& 
d \widetilde{r}^{(N)} _{k+2} (z) - \left( \triangle z+{1 \over z} \right) \widetilde{r}^{(N)} _{k+1} (z) + a \widetilde{r}^{(N)} _{k} (z) = 0,
\end{eqnarray*} 
where $\triangle = \det A = ad - bc$. The characteristic equations with respect to the above recurrences have the same root: if $0 <a, d <1$, then
\begin{eqnarray*} 
\lambda_{\pm} = \lambda_{\pm} (z) = {\triangle z^2 + 1 \mp \sqrt{\triangle^2 z^4 - 2 (ad + bc) z^2 + 1} \over 2dz}. 
\end{eqnarray*}

From now on we consider the $a=d \> (\not= 1/2)$ case with $N=\infty$. We remark that the definition of $\Xi_{1} ^{(\infty)} (n)$ gives $p_1 ^{(\infty)}(n)=0 \> (n \ge 2)$ and $p_1 ^{(\infty)}(1)=1$. So we have $\widetilde{p}^{(\infty)} _1 (z) = z$. Moreover noting $\lim_{k \to \infty} \widetilde{p}^{(\infty)}  _k(z) < \infty$, $0 < \lambda_+ < 1 < \lambda_- \> (0<z<1)$ and using Eq.~(\ref{jyuiti}), we obtain the explicit form 
\begin{eqnarray*}
\widetilde{p}^{(\infty)} _k (z) = z \lambda_+ ^{k-1}, \qquad 
\widetilde{r}^{(\infty)} _k (z) = {\lambda_+ - az \over 1-a} \lambda_+^{k-1},
\end{eqnarray*}
where 
\[
\lambda_\pm 
= {1 + (2a-1)z^2 \mp \sqrt{(2a-1)^2 z^4 -2 \{ a^2+(1-a)^2 \} 
z^2 +1} \over 2az }. 
\]
Therefore we have $\widetilde{p}^{(\infty)} _k (1)= \widetilde{r}^{(\infty)} _k (1) =1$ for any $k \ge 0$. By using these and Lemma \ref{lem9}, we obtain $P^{(\infty)} _{k} (\varphi) = 1$ for $a=d \> (\not= 1/2)$ case. On the other hand, in the case of $a=d=1/2$, the well known result for symmetric classical random walk and Proposition \ref{pro5} (ii) give the same conclusion. So we obtain

\begin{proposition} 
\label{pro10}
Assume that $a=d \in (0,1)$. For any $k \ge 0$ and initial distribution $\varphi = {}^t [\alpha, \beta] \in \Phi$, we have
\begin{eqnarray}  
P^{(\infty)} _{k} (\varphi) = 1.
\label{jyusan}
\end{eqnarray} 
\end{proposition}

Next we consider a finite $N$ case. First we note that $\lambda_+ \lambda_- =1$ since $a=d$. Then $\widetilde{p}^{(N)}_k$ and $\widetilde{r}^{(N)}_k$ satisfy 
\begin{eqnarray}
\widetilde{p}^{(N)}_k(z) = A_z \lambda_+^{k-1} + B_z \lambda_-^{k-1}, 
\qquad
\widetilde{r}^{(N)}_k(z) = C_z \lambda_+^{k-N+1} + D_z \lambda_-^{k-N+1}.
\label{eq:abcd}
\end{eqnarray}
By using Eq.~(\ref{eq:abcd}) and boundary conditions: $\widetilde{p}^{(N)}_1(z)=z$ and $\widetilde{r}^{(N)} _{N-1} (z)=0$, we see that $\widetilde{p}^{(N)} _k (z)$ and $\widetilde{r}^{(N)} _k (z)$ satisfy 
\begin{eqnarray}
&& \widetilde{p}^{(N)}_k(z) = \left( {z \over 2} +E_z \right) \lambda_+^{k-1}
+ \left( {z \over 2} -E_z \right) \lambda_-^{k-1}, 
\label{eq:pp}
\\
&& \widetilde{r}^{(N)}_k(z) = C_z (\lambda_+^{k-N+1}-\lambda_-^{k-N+1}).
\label{eq:rr}
\end{eqnarray}
On the other hand, from Eq.~(\ref{eq:req}) and $\widetilde{r}^{(N)} _{N-1} (z)=0$, we have
\begin{eqnarray}
\widetilde{r}^{(N)}_1(z) 
= z \{ (1-a) \widetilde{p}^{(N)}_2(z) + a \widetilde{r}^{(N)}_2(z) \}, 
\qquad
\widetilde{r}^{(N)}_{N-2}(z) = (1-a) z \widetilde{p}^{(N)}_{N-1} (z).
\label{nomon}
\end{eqnarray}
Combining Eqs.~(\ref{eq:pp}) and (\ref{eq:rr}) with Eq.~(\ref{nomon}) gives 
\begin{eqnarray} 
&& C_z = (1-a) z^2 (\lambda_+^{N-3} - \lambda_-^{N-3}) 
\label{eq:cc}
\\
&& \times \left\{ 
- (\lambda_+^{N-2} -\lambda_-^{N-2})^2
+a z (\lambda_+^{N-2} -\lambda_-^{N-2})
(\lambda_+^{N-3} -\lambda_-^{N-3})
\right.
\nonumber \\
&&
\left.
\qquad \qquad \qquad \qquad \qquad \qquad \qquad \qquad \qquad 
+ (\lambda_+ - \lambda_-)^2 \right\}^{-1}, \nonumber \\
&& 
E_z = - {z \over 2(\lambda_+^{N-2} -\lambda_-^{N-2})}
\bigg[ 2 (\lambda_+ - \lambda_-) 
(\lambda_+^{N-3} -  \lambda_-^{N-3}) 
\label{eq:ee}
\\
&& \times \left\{ 
- (\lambda_+^{N-2} -\lambda_-^{N-2})^2
+a z (\lambda_+^{N-2} -\lambda_-^{N-2})
(\lambda_+^{N-3} -\lambda_-^{N-3})
\right.
\nonumber \\
&& 
\left.
\qquad \qquad \qquad \qquad \qquad \qquad 
+ (\lambda_+ - \lambda_-)^2 \right\}^{-1} + \> (\lambda_+^{N-2} + \lambda_-^{N-2}) \bigg]. 
\nonumber 
\end{eqnarray} 
By using 
\[
\lim_{z \to 1} {\lambda_+^{n} - \lambda_-^{n} \over \lambda_+ - \lambda_-} 
= n,
\]
and Eqs.~(\ref{eq:pp}), (\ref{eq:rr}), (\ref{eq:cc}), (\ref{eq:ee}), we have 
\begin{eqnarray} 
\lim_{z \to 1} \widetilde{p}^{(N)}_k (z) 
&=& 1 - {(1-a)(k-1) \over (1-a)N+2a-1}, 
\label{thatone}
\\
\lim_{z \to 1} \widetilde{r}^{(N)}_k (z) 
&=& {(1-a) \{N-(k+1)\} \over (1-a)N+2a-1},
\label{thattwo}
\end{eqnarray}
for any $1 \le K \le N-1$. On the other hand, the well known result of symmetric classical random walk (that is, $a=d=1/2$) and Proposition \ref{pro5} (ii) imply  
\begin{eqnarray}
P^{(N)} _{k} (\varphi) 
= 1- {k \over N}.
\label{this}
\end{eqnarray}
Therefore, combining  Eqs.~(\ref{thatone}) and (\ref{thattwo}), and Lemma \ref{lem9} for $a \not= 1/2$ with Eq.~(\ref{this}) for $a=1/2$, we obtain 

\begin{theorem} 
\label{th11}
For correlated quantum walk with $a=d$ and $a \in (0,1)$, we have 
\begin{eqnarray*} 
P^{(N)} _{k} (\varphi) 
= {(1-a) (N-k) + (2a-1) \alpha \over (1-a)N+2a-1}, 
\end{eqnarray*} 
for any $1 \le K \le N-1$ and initial distribution $\varphi = {}^t[\alpha, \beta] \in \Phi$.
\end{theorem}

\noindent\par
It should be noted that Theorem \ref{th11} can be also obtained by Theorem 1 in \cite{z}, where $\alpha (=1- \beta) \to a, \> \delta_1= \delta_2 \to 0, \> \rho_1=\rho_2 \to 1, \> \gamma_1 \to 0, \> \gamma_2 \to (1-a)N+2a-1, \> c_1 \to \beta, \> c_2 \to \alpha.$

\par
\
\par\noindent
{\bf Acknowledgments.}  This work is partially financed by the Grant-in-Aid for Scientific Research (B) (No.12440024) of Japan Society of the Promotion of Science. I would like to thank Takahiro Soshi and Kazunori Nakamura for useful discussions.

\par
\
\par\noindent

\begin{small}

\bibliographystyle{plain}

\end{small}

\end{document}